\documentclass[useAMS,usenatbib]{mn2e}

\usepackage[psamsfonts]{amssymb}
\usepackage[dvips]{graphicx}
\usepackage{amsmath,alltt}
\usepackage{multirow}
\usepackage{rotating}

\title[Constraining the Origins of Multiple Star Systems Containing
Merger Products]{An Analytic Technique for Constraining the 
  Dynamical Origins of Multiple Star Systems Containing Merger Products}
\author[Nathan Leigh and Alison Sills]{Nathan Leigh$^{1}$,
  Alison Sills$^{1}$\thanks{E-mail: leighn@mcmaster.ca (NL);
    asills@mcmaster.ca (AS)} \\
$^{1}$Department of Physics and Astronomy, McMaster University, 
1280 Main St. W., Hamilton, ON, L8S 4M1, Canada}
\begin{document}

\pagerange{\pageref{firstpage}--\pageref{lastpage}} \pubyear{2010}

\maketitle

\label{firstpage}

\begin{abstract}
We present a technique to identify the most probable dynamical
formation scenario for an observed binary or triple system containing
one or more merger products or, alternatively, to rule out the
possibility of a dynamical origin.  Our method relies on an analytic
prescription for energy conservation during stellar encounters.  
With this, observations of the multiple star system containing the merger
product(s) can be used to work backwards in order 
to constrain the initial orbital energies of any single, binary or triple
systems that went into the encounter.  The initial semi-major axes of
the orbits provide an 
estimate for the collisional cross section and therefore the
time-scale for the encounter to occur in its host cluster.   

We have applied our analytic prescription to observed binary and
triple systems containing blue stragglers, in particular the triple
system S1082 in M67 and the period distribution of the blue
straggler binaries in NGC 188.  We have shown that both S1082 and 
most of the blue straggler binaries in NGC 188 could have a
dynamical origin, and that encounters involving triples could be a
significant contributor to BS populations in old open clusters.
In general, our results suggest that encounters involving triples
could make up a 
significant fraction of those dynamical interactions that result in stellar
mergers, in particular encounters that produce multiple star systems
containing one or more blue stragglers.  
\end{abstract}


\section{Introduction} \label{intro}

It has been known for some time that encounters, and even direct
collisions, can occur frequently between stars in dense stellar systems
\citep[e.g.]{hills76, hut83a, leonard89}.  In the cores of globular
clusters (GCs), the time between collisions
involving two single stars can 
be much shorter than the cluster lifetime \citep{leonard89}.
The time between encounters involving binary stars can be
considerably shorter still given their much larger cross sections for
collision.  In globular and, especially, open clusters (OCs) with high 
binary fractions, mergers are thought to occur frequently during
resonant interactions involving binaries \citep[e.g.][]{leonard92}.
What's more, collision products have a significant probability of
undergoing more than one 
collision during a given single-binary or binary-binary interaction
since the initial impact is expected to result in shock heating
followed by adiabatic expansion, increasing the cross section for a
second collision to occur \citep[e.g.][]{fregeau04}.    

Several types of stars whose origins remain a mystery are speculated
to be the products of stellar mergers.  Blue stragglers (BSs) in
particular are thought to be produced via the addition of fresh
hydrogen to the cores of low-mass main-sequence (MS) stars.  Recent
evidence has shown that, whatever the dominant BS formation 
mechanism(s) operating in both globular and open clusters, it
is likely to in some way depend on binary stars \citep{knigge09,
  mathieu09}.  The currently favored mechanisms include collisions
during single-binary and binary-binary encounters
\citep[e.g.][]{leonard89}, mass transfer 
between the components of a binary system \citep[e.g.][]{chen08a,
  chen08b} and the coalescence of two 
stars in a close binary due to perturbations from an orbiting
triple companion \citep[e.g.][]{eggleton06, perets09}.

A handful of spectroscopic
studies have revealed that in some GCs there exist BSs with
masses exceeding twice that of the MS turn-off \citep[e.g.][]{shara97,
  knigge08}.  Such massive BSs must have been formed from the
mergers of two or more low-mass MS stars since they
are too massive to have been formed from mass transfer.  In a few 
cases, this can also
be argued for entire BS populations using photometry.  For instance,
\citet{chen08b} performed detailed binary evolution calculations to
study dynamical stability during mass transfer from an evolving
giant star onto a MS 
companion.  Based on their results, it can arguably be inferred that
most BSs in NGC 188 are sufficiently bright that they probably could
not have formed from 
mass transfer alone.  If true, this suggests that most of these BSs
must be the products of stellar meregers.  Regardless of
the dominant BS formation mechanism(s) operating in dense star clusters,
dynamical interactions should play at least some role.  For example,
even if blue stragglers are formed as a result of binary evolution
processes such as mass transfer, the progenitor binaries themselves
should have been affected by at least one dynamical
interaction over the course of their lifetime.

Numerous scattering experiments have been performed to
explore the outcomes of binary-binary and, in particular,
single-binary encounters 
\citep[e.g.][]{mcmillan86, sigurdsson93, fregeau04}.  Most of the
earliest of these studies were 
performed in the point-particle limit, ignoring altogether the often
non-negligible implications of the stars' finite sizes
\citep[e.g.][]{hut83b, mikkola83}.  Later, more realistic
simulations clearly demonstrated the importance of taking into account
the dissipative effects of tidal interactions and direct contact
between stars \citep[e.g.][]{mcmillan87, cleary90}.  As a result of
the increased number 
of free parameters for the encounters and the longer integration times
required to run the simulations to completion, few studies have been
conducted to explore the outcomes of binary-binary encounters or
interactions involving triple systems.

In this paper, we introduce an analytic technique to 
constrain the most probable dynamical origin of an observed binary or
triple system containing one or more merger products.  Provided the
observed system is found within a moderately dense cluster
environment with binary and/or triple fractions of at least a few
percent, the probability is often high that it formed from a merger
during 
an encounter involving one or more binary or triple stars.  In
Section~\ref{method}, we present an equation for energy conservation 
during individual stellar encounters and outline the process for
applying our technique.  Specifically, we present a step-by-step
methodology to evaluate whether or not an assumed dynamical history
could have realistically produced an observed system and describe how
to determine the most probable dynamical formation scenario.  
In Section~\ref{results}, we apply 
our technique to a few observed binary and triple systems thought to
contain merger 
products, in particular a triple system that is thought to contain two BSs and
the peculiar period-eccentricity distribution of the BS binary
population in NGC 188.  We discuss the implications of our results in
Section~\ref{discussion}. 

\section{Method} \label{method}

In this section, we present a general prescription for conservation of
energy during stellar encounters.  We will limit the
discussion to typical interactions thought to occur in globular and
old open clusters, although our technique can be generalized to any
choice of parameter space.  The types of encounters of interest
in this paper will predominantly involve low-mass MS stars with 
relative velocities at infinity ranging from $\lesssim 1$ km/s
to $\sim 10$ km/s \citep[e.g.][]{leonard89, sigurdsson93}.  Our technique
describes how to isolate the most probable dynamical formation history
for an observed binary or triple containing one or more merger products
by providing an estimate for the time required for a given interaction
to occur in a realistic cluster environment.  

We begin by assuming that an observed system was formed
directly from a dynamical interaction (or sequence of
interactions).  In this case, the observed parameters of the system
provide the final distribution of energies for the system resulting 
from the interaction(s).  After choosing an appropriate dynamical
scenario (i.e. whether the objects involved in the interaction(s) are
single, binary or triple stars), we can work backwards using 
energy conservation to constrain the initial
energies going into the encounter.  
This provides an estimate for the initial orbital
energies and therefore semi-major axes of any binaries or 
triples going into the interaction.  This in turn gives the
cross section for collision and hence the time required for the
hypothesized interaction(s) to occur.  

Since the formation event must have happened in the last 
$\tau_{BS}$ years, where $\tau_{BS}$ is the lifetime of the merger
product, a formation scenario is likely only if the derived encounter
time-scale is shorter than the lifetime of the merger product(s).
Conversely, if the derived encounter time-scale is longer than the
lifetime of the merger product(s), then that dynamical formation scenario 
is unlikely to have occurred in the last $\tau_{BS}$ years.  In
general, the shorter the derived encounter time-scale, the
more likely it is that one or more such encounters actually took place
within the lifetime of the merger product(s).  Finally, if the
derived encounter time-scale is longer than $\tau_{BS}$ for every
possible dynamical formation scenario, then a dynamical origin is
altogether unlikely for an observed multiple star system containing
one or more BSs.  Either that, or the encounter time-scales must have
been shorter in the recent past (or, equivalently, the central cluster
density must have been higher). 

\subsection{Conservation of Energy} \label{energy}

Consider an encounter in which at least one of the two bodies involved
is a binary or triple star.  Though a complex exchange of energies occurs,
energy must ultimately be
conserved in any dynamical interaction.  The total energy that goes
into the encounter must therefore be equal to the total energy
contained in the remaining configuration:
\begin{equation}
\begin{gathered}
\label{eqn:energy-conserv}
\sum_i^{N_i} \Omega_i(I_i,\omega_i) + \sum_i^{N_i} W_i(m_i,X_i,Z_i,\tau_i) \\ 
\sum_i^{N_i} U_i(m_i,X_i,Z_i,\tau_i) + \sum_j^{S_i} T_j(M_{cl},M_j) + \\ 
\sum_k^{M_i} \epsilon_k(\mu_k,M_k,a_k) =
    \sum_{ii}^{N_f} \Omega_{ii}(I_{ii},\omega_{ii}) + \\
 \sum_{ii}^{N_f} W_{ii}(m_{ii},X_{ii},Z_{ii},\tau_{ii}) + \sum_{ii}^{N_f} U_{ii}(m_{ii},X_{ii},Z_{ii},\tau_{ii})
 \\
+ \sum_{jj}^{S_f} T_{jj}(M_{jj}) + \sum_{kk}^{M_f} \epsilon_{kk}(\mu_{kk},M_{kk},a_{kk}) +
  \Delta, \\
\end{gathered}
\end{equation}
where $N_i$, $S_i$ and $M_i$ are the total number of stars, objects
(single, binary, triple or even quadruple stars) and orbits,
respectively, that went into the encounter.  Similarly, $N_f$, $S_f$ and
$M_f$ are the total number of stars, objects and orbits remaining after the
encounter. 

We let $\Omega_i = \Omega_i(I_i,\omega_i)$ represent the
bulk rotational kinetic energy in star $i$, which is a function of the star's
moment of inertia $I_i$ and angular rotation rate $\omega_i$:
\begin{equation}
\label{eqn:Omega}
\Omega_i(I_i,\omega_i) = \frac{1}{2}I_i\omega_i^2
\end{equation} 
The moment of inertia is in turn a function of the density profile within a
star, which changes along with its internal structure and composition
as the star evolves.  The moment of inertia is given by \citep{claret89}: 
\begin{equation}
\begin{split}
\label{eqn:mom-inertia}
I &= \frac{8{\pi}}{3}\int_{0}^{R} \rho(r)r^4dr \\
  &= \beta^2mR^2, \\
\end{split}
\end{equation}
where $\beta$ is the radius of gyration.  For example, a typical 1 M$_{\odot}$
star in an old open cluster having an age of $\sim$ 6 Gyrs has $\beta
= 0.241$.  For comparison, a 1 M$_{\odot}$ star in a
typical Milky Way GC having an age of $\sim$ 11 Gyrs has $\beta =
0.357$ \citep{claret89}.  A large spread of rotation speeds have been
observed for MS stars in 
open and globular clusters, with measured values ranging from  $\sim$ 1 -
20 km s$^{-1}$ \citep{mathieu09}.  A 1 M$_{\odot}$
star having a radius of 1 R$_{\odot}$ and a rotation speed of 2 km
s$^{-1}$ with $\beta=0.241$ has $\Omega \sim 10^{36}$ Joules.

We let $W_i = W_i(m_i,X_i,Z_i,\tau_i)$ represent the gravitational
binding energy 
of star $i$, where $m_i$ is the star's mass, $X_i$ is its initial
hydrogen mass fraction, $Z_i$ is its initial metallicity and $\tau_i$
is its age.  For a spherical mass with a density distribution
$\rho_i(r)$, the gravitational binding energy is given by:
\begin{equation}
\begin{split}
\label{eqn:binding}
W_i(m_i,X_i,Z_i,\tau_i) &= \frac{16\pi^2G}{3}\int_0^{R_i}
\rho_i(r)r^4dr \\
                        &= -{\delta}(m_i,X_i,Z_i,\tau_i)G\frac{m_i^2}{R_i}, \\
\end{split}
\end{equation}
where the parameter $\delta$ is chosen to reflect the structure of the
star and is therefore a function of its mass, age and chemical
composition.  For instance, a typical 1 M$_{\odot}$
star with $(X,Z) = (0.70,0.02)$ in an old open cluster with an age
of $\sim$ 6 Gyrs has $\delta = 1.892$.  For
comparison, an older but otherwise identical star in a typical Milky
Way GC with an age of $\sim$ 11 Gyrs has $\delta = 6.337$
\citep{claret89}.  Roughly regardless of age, this gives $|W| \sim 10^{42}$
Joules for a 1 M$_{\odot}$ star with a radius of 1 R$_{\odot}$.

We let $U_i = U_i(m_i,X_i,Z_i,\tau_i)$ represent the total internal
energy contained in star $i$ (i.e. the star's thermal energy
arising from the random motions of its particles).  By solving the
equations of stellar 
structure, the total internal energy of a purely isolated single star
is uniquely determined by its mass, initial composition and age.  
Stars are made up of a more or less virialized fluid so
that, ignoring magnetic fields, the gravitational binding energy of a
star in hydrostatic
equilibrium is about twice its internal thermal energy
\citep{chandrasekhar39}.  Using this version of the virial theorem, a
1 M$_{\odot}$ star having a radius of 1 R$_{\odot}$ has $U \sim 5
\times 10^{41}$ Joules. 

The total translational kinetic energy of object $j$ (single, binary,
triple, etc. star) is represented by $T_j$:
\begin{equation}
\label{eqn:trans}
T_j(M_{cl},M_j) = \frac{1}{2}M_jv_j^2,
\end{equation} 
where $M_j$ is the total mass of the object, 
and $v_j$ is its bulk translational speed.  The
translational velocities of stars in clusters for which energy
equipartition has been achieved as a result of two-body relaxation
obey a Maxwell-Boltzmann distribution, with the heaviest stars
typically having the lowest velocities and vice versa
\citep{spitzer87}.  According to
the virial theorem, the root-mean-square velocity $v_{rms}$ of the
distribution depends on the total mass of the cluster $M_{cl}$.  It 
follows that the velocities of stars in a fully relaxed cluster
are approximately determined by their mass and the total mass of the
cluster.  Assuming that the typical velocity of 
a $<m>$ star is roughly equal to $v_{rms}$,
where $<m>$ is the average stellar mass in the cluster, 
energy equipartition can be invoked in some clusters to approximate
the translational 
kinetic energy, and hence velocity, of a star or binary of mass $M_j$:
\begin{equation}
\label{eqn:v-typ}
v_j = \Big( \frac{<m>}{M_j} \Big)^{1/2}v_{rms}.
\end{equation}
We note that Equation~\ref{eqn:v-typ} can only be applied in clusters
for which the half-mass relaxation time is much shorter than the
cluster lifetime.  At the same time, the tidal truncation of the
velocity distribution must not be significant.  We will return to this
in Section~\ref{general}.  

For a 1 M$_{\odot}$ star with a 
translational speed of 1 km s$^{-1}$ (typical of stars in old open
clusters), we find $T \sim 10^{36}$ Joules.  For comparison, the same
star traveling at a speed of 10 km s$^{-1}$ (typical of stars in
globular clusters) has $T \sim 10^{38}$ Joules.  We can put this into
perspective by equating Equation~\ref{eqn:trans} with
Equation~\ref{eqn:binding}, which shows that
a direct collision between two 1 M$_{\odot}$ stars would require an
impact velocity of $\sim$ 1000 km s$^{-1}$ in order to completely
unbind the merger remnant.

The total orbital energy of orbit $k$ is denoted $\epsilon_k$, 
and is given by:
\begin{equation}
\label{eqn:orbital}
\epsilon_k(\mu_k,M_k,a_k) = -\frac{G{\mu_k}M_k}{2a_k},
\end{equation}
where $\mu_k = m_1m_2/(m_{1}+m_{2})$ is the reduced mass of the orbit, $M_k =
m_1+m_2$ is the total mass and $a_k$ is the orbital semi-major axis.  A
binary composed of two 1 M$_{\odot}$ stars with a period of 1000 days
has $|\epsilon| = 10^{38}$ Joules.  For comparison, an otherwise identical
binary with a period of 1 day has $|\epsilon| = 10^{40}$ Joules.  Since
most stable triples are observed to have outer periods of $\sim$ 1000
days with close inner binaries \citep{tokovinin97, perets09}, it follows
that the total orbital energies of stable triples will be dominated
by the orbital energies of their inner binaries.  

After the encounter occurs, the energies that collectively define the
state of the newly formed system include the rotational kinetic
($\Omega_{ii}(I_{ii},\omega_{ii})$), thermal
($U_{ii}(m_{ii},X_{ii},Z_{ii},\tau_{ii})$) and gravitational binding
($W_{ii}(m_{ii},X_{ii},Z_{ii},\tau_{ii})$) energies of the stars that
are left-over, as well as the total translational kinetic
($T_{jj}(m_{jj})$) and orbital
($\epsilon_{kk}(\mu_{kk},M_{kk},a_{kk})$) energies of any left-over
stars, binaries or triples.  The internal and gravitational binding 
energies of the left-over stars once again depend on their mass,
composition and age.  The mass, composition and evolutionary status of
a star will decide how it responds to tidal interactions (and
therefore how much tidal energy is deposited) since these are the
principal factors that determine its internal structure, in particular
whether or not its envelope is radiative or convective
\citep{podsiadlowski96}.  Finally, we let $\Delta$ represent the
energy lost from the system due to radiation and mass loss.  We do not
expect this term to be significant for the majority of the encounters
we will consider since the rate of mass-loss from low-mass MS stars is
small and the time-scales under consideration are
relatively short compared to the lifetimes of the stars.  Moreover,
the velocity dispersions characteristic of the clusters we will
consider are sufficiently small that we expect most
collisions to have a relatively low impact velocity.  Though
significant mass loss can occur for high impact collisions as often
occur in the Galactic centre, mass loss is only $\lesssim$ 5\% for the low
impact velocity collisions expected to occur in open and globular
clusters \citep{sills01}.  Therefore, we will henceforth assume
$\Delta \sim 0$.

Depending on the initial parameters of the
encounter, one or more terms in Equation~\ref{eqn:energy-conserv} can
often be neglected.  For example, given that the rotational energies
of typical MS stars in 
both open and globular clusters are several orders of magnitude
smaller than the other terms in Equation~\ref{eqn:energy-conserv}, we
can neglect these terms for the types of encounters of interest.  Even
a 1 M$_{\odot}$ star rotating at a rate of 100 km s$^{-1}$ has a
rotational energy of only $\sim 10^{39}$ Joules.  This is considerably 
higher than even the highest rotation rates observed for both BSs and
normal MS stars \citep[e.g.][]{mathieu09}.  From this, we expect
typical rotational energies to be significantly smaller than
the orbital energies of even moderately hard binaries.  
Furthermore, Equation~\ref{eqn:trans} can be combined with
Equation~\ref{eqn:v-typ} and equated to Equation~\ref{eqn:orbital} in
order to define the hard-soft boundary for a given cluster.  If a
binary is hard, its initial orbital energy will outweigh its 
initial translational kinetic energy during typical stellar encounters
\citep{heggie75}.  We do not expect soft binaries to survive for very
long in dynamically-active clusters \citep{heggie75} (i.e. most
binaries are hard), so that 
the translational kinetic energies of the 
stars, binaries and triples typically found in OCs can usually be
neglected when applying Equation~\ref{eqn:energy-conserv}.  This
suggests $T_j = 0$.  However, the final translational kinetic energies
of the stars, 
binaries and triples left-over should not be neglected for encounter
outcomes in which one 
or more stars are ejected with very high velocities.  This can leave
the remaining stars in a much more tightly bound configuration since
stars ejected with high velocities carry off significant amounts of
energy.  Finally, provided there are no very high 
impact collisions and tides dissipate a negligible amount of energy
from the system, we also expect that: 
\begin{equation}
\label{eqn:int-conserv}
\sum_i^{N_i} U_i + \sum_i^{N_i} W_i \sim \sum_{ii}^{N_f} U_{ii} + \sum_{ii}^{N_f} W_{ii}.
\end{equation}

We can simplify our
energy conservation prescription considerably for the majority of
encounters occurring in old OCs.  Neglecting the rotational kinetic
energies of the stars, Equation~\ref{eqn:energy-conserv} becomes: 
\begin{equation}
\begin{gathered}
\label{eqn:energy-conserv2}
\sum_k^{M_i} \epsilon_k(\mu_k,M_k,a_k) = \sum_{kk}^{M_f} \epsilon_{kk}(\mu_{kk},M_{kk},a_{kk}) + \\
\sum_{jj}^{S_f} T_{jj}(M_{jj}) - \Delta_{m}, \\
\end{gathered}
\end{equation}
where we have assumed $T_j = 0$ and $\Delta_{m}$ is the amount of
energy deposited in any merger products formed during the encounter.
In other words, the term $\Delta_{m}$ provides the required 
correction to Equation~\ref{eqn:int-conserv} resulting from 
internal energy being deposited in the merger product(s) as a result
of collisions and tides.  If no mergers occur or all of the orbital
energy of the merging binary is imparted to the other interacting stars,
$\Delta_{m} = 0$. 

Both the total linear and angular momenta must also be conserved
during any dynamical interaction.  This provides two additional
constraints that must also be satisfied, however the total linear and
angular momenta are vector quantities that depend on the angle of
approach as well as the relative orientations of the objects.  As a result,
there are more free parameters to fit when trying to constrain the
initial parameters of an encounter using conservation of linear and
angular momentum.  Therefore, it is considerably more difficult to
extract information pertaining to the initial orbital parameters of
an encounter using conservation of momentum than it is with
conservation of energy. 

We have assumed that exchange interactions do not occur in applying
our technique.  Clearly, this assumption could be invalid for some
systems.  If this is the case, then our method is also invalid.  This
suggests that our technique is ideally suited to clusters for which the
encounter time-scales are comparable to the lifetime of a typical
merger product.  This maximizes the probability that BSs do not experience
any subsequent dynamical interactions after they are formed.  
Of course, most clusters of interest are unlikely to satisfy this
criterion.  
We can assess the probability that an exchange encounter has 
occurred for a particular system by calculating the different
encounter time-scales and comparing to the lifetime of the merger
product.  If the time-scales are sufficiently short, then
the possibility of an exchange interaction having occurred after the
system's formation must be properly addressed.  We will discuss this
further in Section~\ref{discussion}. 

\subsection{Generalized Approach} \label{general}

In this section, we outline a step-by-step
methodology to constrain realistic dynamical formation scenarios
that could have resulted in the production of an observed
stellar system containing one or more merger products.  These steps
are: 

\begin{enumerate}

\item We must first find \textit{qualitative} constraints for the
system's dynamical history and, in so doing, converge on the most
probable formation scenario.  The choice of 
formation history should be guided by the observed properties of the
binary or triple system containing the merger product(s).  The following
guidelines can be applied to find the most probable scenario.

First, the analytic rates for single-single (1+1), single-binary (1+2),
single-triple (1+3), binary-binary (2+2), binary-triple (2+3) and
triple-triple (3+3) encounters can be
compared to obtain a rough guide as to which of these encounter
types will dominate in a given cluster.  The total rate of
encounters of a given type in a
cluster core is well approximated by \citep{leonard89} (see
Appendix~\ref{appendix} for a more generalized form for this
equation):
\begin{equation}
\label{eqn:coll-rate}
\Gamma = N_0n_0\sigma_{gf}(v_{rel,rms})v_{rel,rms},
\end{equation}
where $N_0$ is the number of single, binary or triple stars in the
core and $n_0$ is the mean stellar, binary or triple number density in
the core.  The gravitationally-focused cross section for collision
$\sigma_{gf}$ is given by Equation 6 of \citet{leonard89}.
Gravitationally-focused cross sections for the various encounter types
are provided in Appendix~\ref{appendix} along with the values assumed
for their pericenters.

In general, the number of single, binary and triple stars are 
given by, respectively, $(1-f_b-f_t)N_c$, $f_bN_c$ and $f_tN_c$,
where $N_c = 2/3{\pi}n_0r_c^3$ is the total number of objects in the
core \citep{leonard89}, $f_b$ is the
fraction of objects that are binaries and $f_t$ is the fraction of
objects that are triples.  Assuming for simplicity that 
$v_{rel}$ is roughly equal for all types of encounters, the rates for
two types of encounters can be compared to find the parameter space for
which one type of encounter will dominate over another.  These
relations can be plotted in 
the $f_b-f_t$ plane in order to partition the parameter space for
which each of the various encounter types will occur with the greatest
frequency, as illustrated in Figure~\ref{fig:fb-ft}.  Given a
cluster's binary and triple fractions, this provides a 
simple means of finding the type of encounter that will occur with the
greatest frequency.  Our results
are in rough agreement with those of \citet{sigurdsson93} who found
that single-binary interactions dominate over binary-binary
interactions in clusters having core binary fractions $f_b \lesssim
0.1$, and may dominate for $f_b$ up to 0.25-0.5 in some cases. 

\begin{figure}
\begin{center}
\includegraphics[width=\columnwidth]{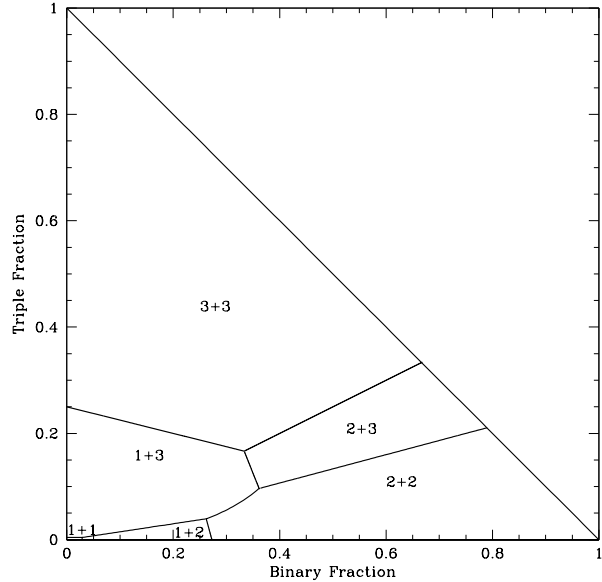}
\end{center}
\caption[Plot showing the parameter space in the f_b-f_t plane for
which each of the various encounter types dominate]{Plot showing
  the parameter space in the $f_b-f_t$ plane for
which each of the various encounter types dominate.  Boundaries
between regions are indicated by solid lines, each segment of which is
obtained by equating two particular encounter rates using
Equation~\ref{eqn:coll-rate} and the relevant cross sections derived
using Equation 6 of \citet{leonard89}.  We assume $a_t =
5a_b$ and $a_b = 90 R$ in obtaining the relations between encounter
rates.  This is a reasonable choice for the ratio between
the average binary and triple geometric cross sections given that
the ratio between the outer and inner orbital semi-major axes of
triples must be relatively large (by a factor of $\gtrsim$ 10) in
order for them to be stable \citep{mardling01}.
\label{fig:fb-ft}}
\end{figure}

Second, the masses of the components of an
observed binary or triple system provide a lower limit for the
number of stars that could have gone into its formation.  The minimum
number of 
stars that must have merged to form a given collision product, labelled
$N_{min,i}$, is equal to the integer nearest to and larger than the
quantity $m_{rem}/m_{TO}$, where $m_{rem}$ is the mass of the merger
remnant and $m_{TO}$ corresponds to the mass of the main-sequence
turn-off (MSTO).  This assumes that $m_{TO}$ has not changed
significantly since the dynamical formation of the system, which
should be valid provided the merger products are significantly more
massive than the turn-off so that their lifetimes are relatively
short.  The number of stars that went 
into an encounter $N_i$ must therefore satisfy:
\begin{equation}
\label{eqn:Nmerge}
N_i \ge \sum_i^M N_{min,i} + M_{0},
\end{equation}
where $M_{0}$ is the number of normal stars (i.e. not formed from
mergers) and $M$ denotes the number of merger products. 

Third, an estimate for the average stellar mass, and therefore the
masses of typical stars expected to undergo encounters, should be
guided by a realistic stellar mass function for 
the host cluster.  First of all, observations have shown that a
significant depletion of 
low-mass stars occurs in dynamically evolved clusters (i.e. those for
which $t_{rh} \ll t_{age}$, where $t_{rh}$ is the half-mass
relaxation time and $t_{age}$ is the cluster age) since they are
preferentially ejected from the cluster during close encounters
\citep[e.g.][]{vonhippel98, bonatto05, demarchi10}.  This suggests
that OCs and the least massive Milky Way GCs should 
have stellar mass functions that, at least in their central cluster
regions, appear eroded at the low-mass end.   From this, we expect
$<m> \lesssim m_{TO}$.  Finally, most of the stars
involved in dynamical encounters should have masses close to or even
slightly greater than the average stellar mass $<m>$.  This is because
gravitational focusing is strongest for the most massive objects,
contributing to a shorter encounter time-scale. 

Finally, the most likely formation scenario will, strictly speaking, 
minimize the total or cumulative encounter time-scale.  However, the
total time-scale required for a dynamical 
formation scenario that involves more than one encounter will
typically be dominated by the second encounter.  This is because,
after an initial encounter has occurred between any two suitable
objects to form a new stellar configuration, the time-scale for a
second encounter to occur is given by the time required for
the product of the initial encounter to experience a
subsequent encounter.  \textit{This increases the encounter time-scale
  by a factor $N_0$.}  In other words, 
Equation~\ref{eqn:coll-rate} provides the time required for
\textit{any} two of the specified objects to experience an
encounter (two binaries, a binary and a triple, etc.).  It can be
multiplied by $N_0$ to obtain an estimate for the time
required for a \textit{specific} object to experience an encounter.
For example, to find the time required for a specific binary to
experience an encounter with another single, binary or triple star, we
must multiply by a factor $f_bN_c$.  It follows that the 
timescale for multiple encounters to occur is considerably longer than
any of the single encounter timescales.  Therefore, unless the number
of either binary or triple stars is very low, 
scenarios involving the fewest number of encounters are generally
preferred since this tends to minimize the total time required for
the encounter(s) to occur in a realistic cluster environment.  

The times \textit{between} the various types of encounters can be
derived using Equation~\ref{eqn:coll-rate} and have been provided in
Appendix~\ref{appendix}.  These time-scales can be multiplied by
$(1-f_b-f_t)N_c$, $f_bN_c$ or $f_tN_c$ to find the time required for a
\textit{particular} single, binary or triple star, respectively, to
encounter another object.  This yields time-scales that are in rough
agreement with Equation 8-125 of \citet{binney87}.  As an example, we
can multiply Equation~\ref{eqn:coll2+2} by $f_bN_c$ to find the time
for a particular binary to experience an encounter with another
binary.  This gives:
\begin{equation}
\begin{gathered}
\label{eqn:tau-2-22}
\tau^b_{2+2} = 2.7 \times 10^{10}f_b^{-1} \Big(\frac{10^3
  pc^{-3}}{n_0}\Big) \\
\Big(\frac{v_{rms}}{5 km/s}\Big)\Big(\frac{0.5
  M_{\odot}}{<m>}\Big)\Big(\frac{1
  AU}{a_{b}}\Big) \mbox{years},
\end{gathered}
\end{equation}

Provided the derived encounter time-scale is shorter than the
lifetime of
the merger product(s), this suggests that the encounter scenario in
question could be realistic and is therefore a candidate formation
history.  Conversely, if the
derived encounter time-scale is longer than the lifetime of
the merger product(s), then the dynamical formation history is unlikely
to have actually occurred.  In general, the shorter the derived
encounter time-scale, the
more likely it is that one or more such encounters actually took place
within the lifetime of the merger product(s). 

The preceding guidelines specify a narrow range of allowed formation
scenarios.  In particular, they constrain the 
number of stars involved in the encounter(s), the types
of objects (i.e. single, binary or triple stars) involved, and the
number of encounters that took place.  In most cases, these guidelines
will converge on a 
single qualitative formation history that is unique up to the possible
initial distribution of energies that describe suitable interactions.

\item Next, we must assign an approximate value based on the
observations to every parameter possible in
Equation~\ref{eqn:energy-conserv}.  Nearly all of the required
information pertaining to the final distribution of energies in
Equation~\ref{eqn:energy-conserv} can be found from spectroscopy
alone, though repeated measurements spread out over a sufficiently
long timeline will typically be required to obtain orbital 
solutions and to detect outer triple companions whenever they are
present.  This gives the final orbital energies $\epsilon_{kk}$ as well
as the gravitational binding energies $W_{ii}$ of the stars according to
Equation~\ref{eqn:orbital} and Equation~\ref{eqn:binding},
respectively.  Since merger products have been shown to typically be
in hydrodynamic  
equilibrium \citep[e.g.][]{sills01}, the stars' internal energies
$U_{ii}$ can then be approximated using the virial theorem.  Alternatively,
stellar models can be used together with photometry.  

The measured broadening of spectral lines gives an estimate for the
stars' rotation speeds (although there is a strong dependence on the
angle of inclination of the stars' axis of rotation relative to the
line of sight), which in turn provides their rotational 
kinetic energies $\Omega_{ii}$ according to Equation~\ref{eqn:Omega}.  In
conjunction with proper motions, radial 
velocity measurements also provide an estimate for the systemic
velocity of the final stellar configuration relative to the
cluster mean, which in turn gives its translational kinetic energy $T_{jj}$
according to Equation~\ref{eqn:trans}.  
Finally, the cluster velocity dispersion provides an estimate of the
relative velocity at infinity for a typical encounter, which in turn
decides the initial translational kinetic 
energies $T_j$ of stars or binaries involved in the encounter.  Under
the assumption of energy equipartition for both single and binary
stars, the initial velocities of the impactors can be approximated by
Equation~\ref{eqn:v-typ}.  The assumption of energy equipartition
should be valid in clusters for which the half-mass relaxation time is
considerably shorter than the cluster lifetime, and this is the case
for most GCs \citep{harris96, deangeli05}.  On the other hand, this
assumption is likely invalid for most open clusters.  This is because
the tidal truncation of the velocity distributions are significant in
OCs since they are much less centrally concentrated.  In this case,
the initial velocities of the impactors can be approximated from the
velocity dispersion, which is nearly independent of mass.

\item We can now obtain \textit{quantitative} constraints for the initial
encounter(s).  Once we have decided on a qualitative encounter
scenario that could have produced the observed merger product(s), we
can estimate the orbital energies of the initial binaries 
or triples going into the encounter using
Equation~\ref{eqn:energy-conserv2}.  This gives us an equation
that relates the initial orbital semi-major axes of all orbits going
into the encounter (to each other).  If only one orbit goes into the
encounter, then we 
can solve for it explicitly.  From this, we can constrain the initial
collisional cross sections for realistic encounters.  If the derived 
cross section is significantly smaller than the average semi-major
axes of all binaries and/or triples in the cluster, then
we can infer that the time required for the encounter to occur is
significantly longer than the corresponding time-scale given in
Appendix~\ref{appendix}.  If we use our derived
cross section found from 
Equation~\ref{eqn:energy-conserv2} instead of this average semi-major
axis, this should give us an idea of the time required for that 
particular type of encounter to occur.  Strictly speaking, however,
this is only a rough approximation since these are typical time-scales
found using the average period (or, equivalently, semi-major axis and
hence cross section). 

If we find that the derived encounter time-scale is 
longer than the lifetime of a typical merger product, then the chosen
formation history is unlikely to have actually occurred.  In this
case, it becomes necessary to 
re-evaluate the possible dynamical formation histories of the observed
binary or triple system, choosing the
next most likely qualitative scenario for further quantitative
analysis.  These
steps can be repeated until either a suitable formation history is
found or the list of possibilities is exhausted so that the only
remaining conclusion is that the observed binary or triple system is
unlikely to have a dynamical origin.

\end{enumerate}

\section{Results} \label{results}

In this section, we apply our technique to two particular
cases of observed binaries and triples containing merger products.  The 
first is an observed triple system in the old open cluster M67 that is
thought to contain two BSs \citep{vandenberg01,
  sandquist03}.  The second is the 
period-eccentricity distribution of the BS binary
population of the OC NGC 188, which bears a remarkable
resemblance to M67 \citep{mathieu09}.  After determining the most
probable qualitative formation 
scenarios, we obtain quantitative constraints for suitable initial
conditions that could have produced the observed orbital parameters.

\subsection{The Case of S1082} \label{s1082}

S1082 is believed to be a triple system in the old OC M67
\citep{vandenberg01, sandquist03}.  The observations suggest that a
distant triple companion orbits a close binary containing a BS and
another peculiar star.  The companion to the BS has a photometric
appearance that puts it close to the MSTO in the CMD and yet,
curiously, its derived mass is significantly greater than that of the
turn-off.  The outer companion is a BS in its own right, so that S1082
is thought to be composed of two BSs.  Although both the 
inner and outer components of this suspected triple have systemic
velocites that suggest they are both cluster members, it is important
to note that there is no direct evidence proving a dynamical link
between the two \citep{sandquist03}.  Assuming for the time being that
a dynamical link does exist, we can apply the procedure outlined in
Section~\ref{general} to the case of S1082:

\begin{enumerate}

\item Before applying our technique, it is important to convince
  ourselves that a dynamical origin is possible for the observed
  system.  This is certainly the case for S1082 since no known BS
  formation mechanism could have produced the observed stellar configuration
  without at least some help from dynamical interactions.  The first
  step of our procedure is to find qualitative constraints and,
  in so doing, isolate the most probable encounter scenario.  

First, we need to know the cluster binary and triple fractions in
order to use Figure~\ref{fig:fb-ft} to find the encounter type
occurring with the greatest frequency.  From this, we find that 2+2
encounters presently dominate in M67.  \citet{fan96} showed
that observations of M67 are consistent with a 
cluster multiple star fraction $\sim 50\%$.  More recent
studies report a lower limit for $f_b + f_t$ that is consistent with their
results \citep[e.g.][]{latham05, davenport10}.  Radial
velocity surveys and simulations
of dynamical interactions suggest that old OCs like
M67 are likely to host a number of triples with outer periods
$\lesssim$ 4000
days \citep[e.g.][]{latham05, ivanova08}.  We assume 
$f_t/f_b \sim 0.1$ and, using the result of \citet{fan96}, this gives
$f_b \sim 0.45$ and $f_t \sim 0.05$.  Our assumed ratio $f_t/f_b$ is
slightly lower than found for the field, or $f_t/f_b \sim 0.2$
\citep{eggleton08}. 

Second, we need to constrain the number of stars that went into the
encounter.  The mass of the MSTO in M67 is estimated to be $\sim$ 1.3
M$_{\odot}$ \citep{mathieu09}.  The total mass of S1082 is $\sim$ 5.8
M$_{\odot}$ \citep{sandquist03}.  From Equation~\ref{eqn:Nmerge}, its
formation must therefore have involved at least 5 stars.  

At this point, we can conclude that a single 2+2 encounter could not
have produced S1082 since this scenario involves only 4 stars and we
know that at least 5 stars are needed.  We must therefore consider
either a single 2+3 or 3+3 encounter, or a multiple encounter
scenario.  In order to isolate the most probable of these
possibilities, we must calculate and compare their encounter
time-scales.  To do this, we require
estimates from the observations for a few additional 
cluster parameters.  For M67, the core radius is $r_c$ $\sim$ 1.23
pc \citep{bonatto05, giersz08}.  From this and the central velocity
dispersion, we can calculate the central mass density using Equation
4-124b of
\citet{binney87}, which gives $\rho_0$ $\sim$ $10^{1.9}$ M$_{\odot}$
pc$^{-3}$.  The central stellar number density can then be
approximated according to:
\begin{equation}
\label{eqn:num-density}
n_0 = \frac{\rho_0}{<m>}\frac{M}{L},
\end{equation}
where M/L is the cluster mass-to-light ratio and should be
around 1.5 for an OC as old as M67 \citep[e.g.][]{degrijs08}.  We take
$f_t \sim 0.05$ and assume most stable triples have 
outer periods of $P \sim 1000$ days so that $a_t \sim 3$ AU (assuming
all three stars have a mass of 1 M$_{\odot}$).  We also take $f_b \sim
0.45$ and assume an average binary period of $P \sim 100$ days so that
$a_b \sim 0.6$ AU (assuming both components have a mass of 1
M$_{\odot}$).  We assume an average stellar mass of $<m> \sim 
1.0$ M$_{\odot}$, which is in reasonable agreement with
the observations \citep{girard89}.  Assuming
that the average mass of merger remnants is equal to $2<m>$
and extrapolating the results of
\citet{sills01} for solar metallicity and more massive parent stars,
we will assume that the typical lifetime of a merger product is
$\tau_{BS} \sim 1.5$ Gyrs.

A comparison of the relevant encounter time-scales suggests that the
most probable dynamical formation scenario for S1082 is a single 2+3 
encounter, although a single 3+3 encounter is almost equally as
probable.  Given our assumptions, we find 
$8.9 \times 10^8$ years and $3.3 \times 10^9$ years for the times between
2+3 and 3+3 encounters, respectively, in the core of M67.  From this,
we expect approximately 
two and zero 
2+3 and 3+3 encounters, respectively, to have occurred within the last
$\tau_{BS}$ years.  

From Equation~\ref{eqn:tau-2-22}, we find that the time for a particular
binary to encounter another binary is $7.3 \times 10^{10}$ years.
Similarly, the time for a particular quadruple to encounter another
binary is $8.9 \times 10^9$ years (using Equation~\ref{eqn:tau-2-22}
and assuming the quadruple has a mass
of $4<m>$ and its geometric cross section is twice as large as the
average outer semi-major axis of triples).  Since these time-scales
are considerably longer than the
cluster lifetime, this suggests that a formation scenario for S1082
involving back-to-back 2+2 encounters is unlikely, even if the second
encounter occurred sufficiently soon after the first that all four
stars comprising the initial pair of interacting binaries are still
gravitationally bound.  If we replace one of the 3 binaries involved
in this scenario with a triple system, the total encounter time
remains longer than the cluster lifetime.  

\item Before more quantitative constraints can be found, we
  must refer to the literature in order to obtain estimates 
for every term in Equation~\ref{eqn:energy-conserv2}.  The
observations suggest that a binary system composed of a 2.52
M$_{\odot}$ blue straggler (component 
Aa) and a 1.58 M$_{\odot}$ MS star (component Ab) has a period of $P
\sim 1.068$ days \citep{vandenberg01}.  There is also evidence for a
1.7 M$_{\odot}$
blue straggler companion (component B) that acts as a stable outer
triple with a period of $P \sim 1188.5$ days \citep{sandquist03}.
From this, we can calculate the final orbital energies ($\epsilon_A$
and $\epsilon_B$) of the triple:
\begin{equation}
\label{eqn:orb-f}
|\epsilon_A| + |\epsilon_B| \sim |\epsilon_A| \sim 10^{41} J,
\end{equation}
where we have used the fact that $|\epsilon_A| \gg |\epsilon_B|$ since
$|\epsilon_B| \sim 10^{39}$ J.  
It is important to note that these peculiar stars are often
underluminous, so the inferred mass of the tertiary companion should
be taken with caution \citep{vandenberg01}.  

The systemic radial velocity of the system is $33.76 \pm 0.12$ km/s
\citep{sandquist03}.  Although this only provides us with an estimate
for the systemic velocity of S1082 in one dimension, it is consistent
with the cluster mean 
velocity of 33.5 km/s.  Therefore, the available knowledge for the
systemic velocity of S1082 is consistent with its final
translational kinetic energy being negligible.  From this, we take
$T_{A,B} \sim 0$.  The central 
velocity dispersion in M67 is only 0.5 km s$^{-1}$ \citep{mathieu86,
  mathieu90}.  This suggests that the relative velocity at infinity, 
and therefore the typical impact velocities of collisions, should be
small compared to the significant orbital energy of component A in
S1082 (and any very hard binaries that went into the encounter).  From
this, we take $\Delta_m \sim 0$.

\item We are now ready to obtain quantitative constraints for the
formation of S1082.  We can use Equation~\ref{eqn:energy-conserv2} 
since M67 is an old OC with a small central velocity
dispersion so that the assumptions used to derive this equation are valid.  
At this point, we must consider the details of our chosen formation
scenario more carefully in order to choose a set of initial conditions
that will satisfy Equation~\ref{eqn:energy-conserv2} as well as
reproduce the observed parameters of S1082.  In doing so, we find that
it is not possible to form S1082 with a single 2+3 encounter.  This is
because 5 stars with masses $m < m_{TO}$ are insufficient to form both
an inner binary with a total 
mass $\sim 4.1$ M$_{\odot}$ and an outer tertiary companion with a
mass 1.7 M$_{\odot}$.  That is, the total mass of the inner binary is
larger than three times the mass of the MSTO so that its formation
must have involved four or more stars.

A single 3+3 encounter is therefore the most probable formation
scenario for S1082.  Although there are a number of ways we can
reproduce the observed parameters of S1082 with a 3+3 encounter,
including the component masses, we must use 
Equation~\ref{eqn:energy-conserv2} to identify the most probable of
these scenarios.  To do this, we can solve for 
the initial orbital energies of all binaries and/or triples going
into the encounter, which will in turn constrain their
initial semi-major axes and therefore cross sections for collision.  
In applying Equation~\ref{eqn:energy-conserv2}, we are only concerned
with the initial and final orbital energies of the inner binaries of
the triples.  This is because the orbital energies of any outer triple
companions will be significantly outweighed by the orbital energies of
their inner binaries.  It follows that the contribution from the outer
orbital energies of triples to the total energy of the encounter will
be negligible. 

From Equation~\ref{eqn:energy-conserv2}, we find 
that the formation of S1082 should have involved at least one hard
binary with 
$|\epsilon_i| \sim 10^{41} J$ whose components did not merge (with each
other) during the encounter in order to account for the significant
orbital energy of component A.  
This need not be the case, however, if one or more stars were ejected
from the system with an escape velocity of $\gtrsim 100$ km/s.  Of
course, this would require an increase in the total number of
stars involved in the interaction and therefore a single encounter
scenario involving 7 or more stars, and hence one or more quadruple
systems.  Although very few observational constraints for the fraction
of quadruple systems in clusters exist, this seems unlikely.

The orbital energies of the inner binaries of the two triples
initially going into the encounter should both be on the order of
$10^{41}$ J.  Although we have found
that only one hard binary with $|\epsilon_i| \sim 
10^{41}$ J is required, triples are only dynamically stable if the
ratio of their inner to outer orbital periods is large
(roughly a factor of ten or more) \citep[e.g.][]{mardling01}.  We do
not expect very wide binaries to survive for very long in dense
cluster environments, which suggests that the inner binary of every
triple is very hard.  We expect from this that these inner binaries
should, to within an order of magnitude, all have roughly the same
orbital energy.  

The presence of outer triple companions is required in order for most
very hard binaries to actually experience encounters within the
lifetime of a typical merger product.  Assuming masses of 1
M$_{\odot}$ for both components, an orbital energy of $10^{41}$ J
corresponds to a period $\sim 0.4$ days, or a semi-major axis $\sim
0.02$ AU.  Therefore, the cross section for collision for a 2+2 encounter
in which both binaries are very hard is smaller than the average
cross section for a 2+2 encounter (found from the observed binary
period distribution) by a factor of $\sim 30$.  This suggests that the
time required for an 
encounter to occur between two very hard binaries is considerably
longer than the cluster lifetime.  This is not the case if the hard
binaries have triple companions, however, since the outer orbit
significantly increases the collisional cross section and hence
decreases the encounter time-scale.  

Energy conservation also suggests that if S1082 did form from a single 3+3
encounter, it is likely that the close inner binaries of the two triples
collided directly.  
If two separate mergers then subsequently occurred during this
interaction, this could have reproduced the observed orbital
parameters 
of the close inner binary of S1082 (component A).  The formation of
the outer triple companion is more difficult to explain via a
single 3+3 encounter since it also appears to be the product of a
stellar merger.  Nonetheless, if the outer companions of both triples
undergoing the encounter end up orbiting the interacting (or merged) 
pair of close inner binaries at comparable distances, it is possible
that their orbits would overlap as a result of gravitational focussing.
Although this seems unlikely, it could produce a blue straggler of the
right mass and orbital period to account for component B.  We 
will return to this possibility in Section~\ref{discussion}. 

\end{enumerate}

Even though the analytic estimates presented here are only approximate,
they serve to 
highlight the low probability of a system such as S1082 having formed
dynamically in M67.  Based on our results, the most likely formation
scenario for S1082 is 
a single 3+3 encounter, although we expect very few, if any, to have
occurred in the lifetime of a typical merger product.  This need not
be the case, however, provided M67 had a 
higher central density in the recent past since this would increase
the total encounter frequency.  This will be discussed further in
Section~\ref{discussion}.  

\subsection{The Period-Eccentricity Distribution of the Blue Straggler
  Binary Population in NGC 188}

\citet{mathieu09} found 21 blue stragglers in the old open cluster NGC
188.  Of these, 16 are known to have a binary companion.  Orbital
solutions have been found for 15 of these known BS binaries.  From
this, \citet{mathieu09} showed that the BS binary population in NGC
188 has a curious 
period-eccentricity distribution, with all but 3 having periods of
$\sim 1000$ days.  Of these three, two have periods of $\lesssim 10$
days (binaries 5078 and 7782).  Interestingly, one of these
short-period BS binaries has a non-zero
eccentricity.  The normal MS binary population, on the other hand, shows
no sign of a period gap for $10 \lesssim P \lesssim 1000$ days
\citep{mathieu09}.  We can apply the procedure outlined in
Section~\ref{general} to better understand how we expect mergers
formed during dynamical encounters to contribute to the BS binary
population in NGC 188.  Although the method described in
Section~\ref{general} treats one system at a time, we will apply our
technique to the BS binary population of NGC 188 as a whole.

\begin{enumerate}

\item Before applying our technique, we must satisfy ourselves
  that a dynamical origin is possible for a large fraction of the
  observed BS population.  
  Several examples of evidence in favour of a dynamical origin exist.
  For one thing, most BSs in NGC 188 have been
  found to have binary companions \citep{mathieu09}.  This should not
  be the case if most BSs are the products of the coalescence of
  isolated binary systems.  On the other hand, a binary companion
  should be expected if the BSs were formed from mass transfer.  However,
  the results of \citet{chen08b} could suggest that most of the BSs in
  NGC 188 were not formed from mass transfer alone, providing indirect
  evidence that many BSs were formed from mergers.  A binary companion
  should also be expected if the coalesced binary had a tertiary 
  companion to begin with.  Although the Kozai mechanism can act to decrease the
  orbital separation of the inner binary of a triple, an additional
  perturbation is often required in order to fully induce the binary
  to merge \citep{perets09}.  All of this suggests that
  dynamical interactions likely played at least some role in the formation
  of a significant fraction of the BS binary population in NGC 188.

The first step of our technique is to find the
most commonly occurring type(s) of encounter(s).  \citet{geller08}
found a completeness-corrected multiple star fraction $f_b + f_t
\sim 0.27$ out to a period of $\sim 4000$ days.  This represents a
lower limit since it does not include binaries with $P \gtrsim 4000$
days.  Using the same lower limit for the ratio
$f_t/f_b \sim 0.1$ found by \citet{latham05}, Figure~\ref{fig:fb-ft}
suggests that 1+2 encounters should currently dominate in NGC 188.

Second, we must constrain the minimum number of stars required to form
the observed systems.  
From this, we can show that a merger occurring during all but a 1+1
encounter could produce a BS in a binary since this is the only type
of encounter that involves less than 3 stars.  By dividing the total
mass of an observed 
system by $m_{TO}$, we can find an estimate for the minimum number of
stars required to have been involved in its formation.  Since all but
one BS binary contain only a single merger product with a mass $\ge
m_{TO}$, realistic dynamical formation scenarios for these systems 
require 3 or more stars.  Binary 7782, on the other hand, is thought
to contain two BSs so that if its formation involved two separate
mergers it must have required 4 or more stars.

A more quantitative comparison of the different encounter types is
required since we do not yet know if \textit{enough} encounters
occurred in the last $\tau_{BS}$ years to account for all 16 observed
BS binaries.  To do this, we require estimates from the observations
for a few additional 
cluster parameters in order to calculate and compare the relative
encounter time-scales.  NGC 188 has been found to
have a central density of $\rho_0 = 10^{2.2}$ 
L$_{\odot}$ pc$^{-3}$ \citep{sandquist03} and a core radius of $r_c \sim
1.3$ pc \citep{bonatto05}.  
We found an average stellar mass for the cluster of $<m> \sim 0.9$
M$_{\odot}$.  This was done by determining the cluster luminosity
function using only those stars known to be cluster members from the
proper-motion study of \citet{platais03}.  With this, a theoretical
isochrone taken from \citet{pols98} was used to determine the cluster
mass function and the average stellar mass was calculated.  

A sufficient number of suitable dynamical interactions 
should have occurred in the last $\tau_{BS}$ years for the formation
of every BS binary in NGC 188 to have been directly mediated by the
cluster dynamics.  Assuming once again that $\tau_{BS} \sim 1.5$ Gyrs,
the encounter time-scales derived in Appendix~\ref{appendix} suggest
that a minimum of nine, eight, eight, three and one 1+2, 2+2, 1+3,
2+3 and 3+3 encounters, respectively, 
occurred within the lifetime of a typical merger product.  It follows
that at least $\sim 29$ dynamical encounters
should have occurred in NGC 188 in the last $\tau_{BS}$ years.  Of
these 29 encounters, 12 should have involved triples.  In deriving these time-scales,
we have taken $f_b + f_t \sim 0.27$ from \citet{geller09} and have
adopted the same ratio $f_t/f_b \sim 0.1$ as found for M67.
Consequently, these represent lower 
limits for $f_b$ and $f_t$, so that our derived encounter rates are
also lower limits.  
We have also assumed an 
average outer semi-major axis for triples of $a_t \sim 3$ AU 
(corresponding to a period of $\sim$ 1000 days for a binary composed
of two 1 M$_{\odot}$ stars).

\item The next step is to apply our energy conservation prescription
  to the observed BS population.  This will allow us to constrain the orbital
  energies of typical binaries or triples expected to form BS binaries
  via dynamical interactions.  First, we must refer to the literature
  in order to obtain estimates for every term in
  Equation~\ref{eqn:energy-conserv2}. 
From the observed BS period-eccentricity distribution in NGC 188, we know
that most BS binaries have periods of $\sim 1000$ days (we will call
these long-period binaries), although there
are a couple with very short periods of $\sim 10$ days (which we will
call short-period binaries).  This provides 
an estimate for the final orbital energies $\epsilon_{kk}$ of BS binaries
formed during dynamical interactions.  Specifically, we find
$|\epsilon_{kk}| \sim 10^{39}$ J and $|\epsilon_{kk}| \sim 10^{40}$ J for the
final absolute orbital energies of the long- and short-period BS
binaries, respectively.

Every BS in NGC 188 with a high cluster membership probability
has both radial and proper motion velocities that, to within their
respective uncertainties, are consistent with the observed central
velocity dispersion of $\sigma_0 = 0.41$ km s$^{-1}$ \citep{platais03,
  geller08, geller09}.  From this, we assume that the
final translational kinetic energies of any binaries or triples formed
from dynamical interactions will be negligible, or $T_{jj} \sim 0$.  As
for M67, we assume $\Delta_m 
\sim 0$ since we expect low impact velocities for collisions as a
result of the low velocity dispersion in NGC 188.

\item We are now equipped with estimates from the literature that will
  allow us to obtain quantitative constraints for specific encounter
  scenarios.  In particular, we can use
  Equation~\ref{eqn:energy-conserv2} to constrain the initial 
  orbital energies of all binaries and/or triples going into an 
  encounter.  We can also constrain the specific details of
  interactions for which the encounter outcome reproduces the observed
  parameters of the BS binary.  We will consider 
  energy conservation separately for two different classes of BS
  binaries, namely short- and long-period. 

The short-period BS binaries have large (absolute) orbital energies.
Equation~\ref{eqn:energy-conserv2} suggests that this energy requires
that at least one hard binary was involved in the encounter.
Alternatively, an encounter which involved only softer binaries must
have resulted in at least one star escaping with a significant
velocity ($\gtrsim 100$ km/s).  However, in order to produce a merger
product in a short-period binary, encounters involving triples
are the most favoured.  Binary-binary (and especially 1+2) encounters
which involve at least one hard binary will have smaller
cross sections for collision than those encounters involving wide
binaries.  But energy conservation suggests that 2+2 encounters
involving wide binaries are unlikely to produce a merger product in a close
binary.  Stable triples, on the other hand, contain a hard inner
binary \citep{mardling01} which will naturally account for the large
orbital energy of 
the resulting BS binary.  Stable triples also have a large cross
section for collision because of the wide orbit of the tertiary
companion.  The times between 1+2, 2+2 and 1+3 
encounters are all comparable, suggesting that most encounters
involving hard binaries are 1+3 encounters. 

The short-period BS binaries could have formed from a
direct stellar collision that occurred within a dynamical
encounter of a hard binary and another single or (hard) binary star
(we call this Mechanism I).  If at least one of the objects going into
the encounter was a triple, then four or more stars were involved in
the interaction.  Therefore, if binaries 5078 
and 7782 were formed from this mechanism, they could 
possess triple companions with sufficiently long periods that they
would have thus far evaded detection.  This is consistent with the
requirements for both conservation of energy and angular momentum.  
Interestingly, the presence of an outer triple companion could also
contribute to hardening these BS 
binaries via Kozai cycles operating in conjunction with tidal friction
\citep{fabrycky07}.  In this case, Equation~\ref{eqn:energy-conserv}
shows that tides can contribute to making a binary's orbital energy
more negative by depositing internal energy into the component stars.

Although 2+2 encounters should also contribute to the observed BS
population, most of these should occur between a short-period binary
and a long-period binary.  This is because most encounters resulting in mergers
involve very hard binaries and binaries with long periods have 
large cross sections for collision and therefore short encounter
times.  Equation~\ref{eqn:energy-conserv2} indicates that
the minimum period of a BS binary formed during a 2+2 encounter is
usually determined by the orbital energy of the softest binary going
into the encounter.  This is because the most likely merger scenario
is one for which the hard binary is driven to coalesce by imparting
energy and angular momentum to other stars involved in the interaction 
\citep{fregeau04}.  Assuming most of 
this energy is imparted to only one of the stars causing it to escape 
the system, we can take $\Delta_m$ in
Equation~\ref{eqn:energy-conserv2} to be very small and the orbital
energy of the left-over BS 
binary will be comparable to the orbital energy of the initial wide
binary going into the encounter.  As more energy is imparted to the
star left bound to the 
merger product over the course of the interaction, its final orbital
separation effectively increases.  This in turn contributes to an increase 
in the final orbital energy of the left-over BS binary.  
Finally, from Equation~\ref{eqn:coll2+2}, the period
of the initial wide binary should be relatively long (roughly $\gtrsim
1000$ days) since the contribution from the very hard binary to the
total cross section for collision is negligible.  Otherwise, the time
required for such a 2+2 encounter to occur could exceed $\tau_{BS}$.

Now let us consider the long-period BS binaries in NGC 188.  Based on 
Equation~\ref{eqn:energy-conserv2}, we expect encounters involving
3 or more stars and only one very hard binary to typically produce
long-period BS binaries if the hard binary is driven to merge during
the encounter.  This is because the hard binary merges so that its
significant (negative) orbital energy can only be re-distributed to
the other stars by giving them positive energy.  Since the only other
orbits involved in the interaction are wide, the left-over BS binary
should also have a wide orbit.  Alternatively, BSs formed during
interactions involving more than one very hard 
binary should be left in a wide binary provided enough energy is
extracted from the orbit of the binary that merges.  In this case, a
significant 
fraction of this energy must be imparted to the other stars in order
to counter-balance the significant orbital energy of the other very
hard binary.  If not, the other short-period 
binaries are required to either merge or be ejected from the
system.  Otherwise
Equation~\ref{eqn:energy-conserv2} suggests that the left-over BS
binary should be very hard.  However, it is important to recall that we
are neglecting other non-dynamical mechanisms for energy extraction.
We will return to this last point in Section~\ref{discussion}.  

In order to help us obtain more quantitative estimates for the
long-period BS binaries, consider two additional
mechanisms for mergers during dynamical interactions that involve both
short- and long-period orbits.  First, a
merger can occur if a sufficient amount of orbital energy is extracted
from a hard binary by other interacting stars (called Mechanism IIa).
Alternatively, a merger can occur if the encounter progresses in such
a way that the eccentricity of a hard binary becomes sufficiently
increased that the stellar radii overlap, causing the stars to
collide and merge (called Mechanism IIb).  In the case of Mechanism
IIb, most of the orbital energy of the close binary must end 
up in the form of internal and gravitational binding energy in the
merger remnant after the majority of its orbital angular momentum has been
redistributed to the other stars (and tides have extracted orbital
energy).  In the case of Mechanism IIa, however, most 
of the orbital energy of the close binary must be imparted to
one or more of the other interacting stars in the form of bulk kinetic
motion.  Consequently, one or more stars are likely to obtain a
positive total energy and escape the system.  This need not
necessarily be the case for 2+3 and 3+3 encounters provided the second
hardest binary orbit involved in the interaction has a sufficiently
negative energy.  

Regardless of the type of encounter,
Equation~\ref{eqn:energy-conserv2} shows that the extraction of
orbital energy from a hard binary in stimulating it to merge should 
increase the final orbital period of a BS binary.  To illustrate this,
we will consider 1+3 encounters since the predictions of energy
conservation are nearly identical for the other encounter types of 
interest.  Moreoever, we have shown that encounters with triples
are the most likely to involve very hard binaries.
Equation~\ref{eqn:energy-conserv2} can be re-written for 1+3
encounters:  
\begin{equation}
\label{eqn:energy-conserv3}
\epsilon_{12,4} = \epsilon_{12,3} - f_{12} \times \epsilon_{12},
\end{equation}
where we have assumed that stars 1 and 2 comprise the initial hard
inner binary of the triple, star 3 is the initial outer triple
companion and star 4 is the interloping single star.  Stars 1 and 2
are assumed to merge during the encounter by exchanging energy and
angular momentum with stars 3 and 4.  We further assume that enough
energy is imparted to star 3 that it escapes the system.  We let $f_{12}$ 
represent the fraction of energy extracted from the orbital energy of
the hard inner binary of the triple in the form of bulk kinetic motion
by star 4.  Since the remaining orbital energy
of the close inner 
binary will end up in the form of internal and gravitational binding
energy in the merger remnant, we have assumed $(1 + f_{12}) \times
\epsilon_{12} \sim T_3 - \Delta_m$ in obtaining
Equation~\ref{eqn:energy-conserv3} from
Equation~\ref{eqn:energy-conserv2}.

This formation mechanism could leave the BS as a single star if 
$f_{12}$ exceeds a few percent.  As shown in
Figure~\ref{fig:sintri}, the period of a BS binary formed 
during a 1+3 encounter via Mechanism IIa ($P_{12,4}$) is only slightly
smaller than the period of the outer orbit of the triple initially
going into the encounter ($P_{12,3}$).  This assumes, however, that no
energy is exchanged between the hard inner binary and star 4
(i.e. $f_{12} = 0$).  The
predictions from energy conservation for this case are therefore
identical for Mechanism IIb.  In
general, as the amount of energy extracted from the hard inner binary
of the triple by star 4 increases, so too does the rate at
which $P_{12,4}$ increases with increasing $P_{12,3}$.  
If the amount of energy extracted is $\gtrsim$ 5\% of the orbital
energy of the initial inner binary of the triple, $P_{12,4}$ becomes a
very steeply increasing function of $P_{12,3}$.  

\begin{figure}
\begin{center}
\includegraphics[width=\columnwidth]{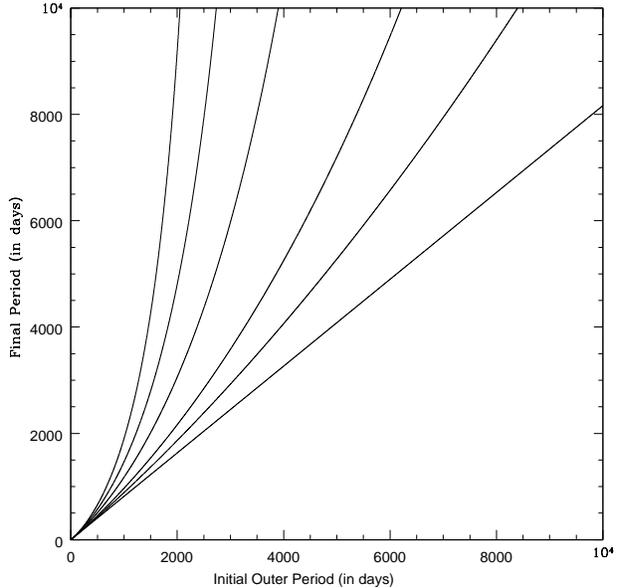}
\end{center}
\caption[Plot showing the typical periods of BS binaries expected to
form during 1+3 encounters in which the hard inner binary of the
triple merges]{Plot showing the typical periods of BS binaries
  formed from 1+3 encounters in which the hard inner binary of the
  triple merges.  As described in the text, the period of the BS binary
  formed during the interaction is denoted $P_{12,4}$ and corresponds
  to the y-axis, whereas the period of the outer orbit of the triple
  initially going into the encounter is denoted $P_{12,3}$ and
  corresponds to the x-axis.  The lower straight line 
  corresponds to the case where no energy was extracted from the inner
  binary of the triple by star 4 (i.e. $f_{12}=0$).  As the amount of
  energy extracted increases, however, so too will the 
  rate at which $P_{12,4}$ increases with increasing $P_{12,3}$.
  Cases where $f_{12}=0.005$, $f_{12}=0.01$, $f_{12}=0.02$,
  $f_{12}=0.03$ and $f_{12}=0.04$ are shown as lines of increasing
  slope.
\label{fig:sintri}}
\end{figure}

Interestingly, the two general qualitative merger scenarios described above 
(Mechanisms I and II) naturally create a bi-modal period distribution
similar to the period gap observed for the BS binaries if we assume
that 1+3 encounters produced these objects.  To illustrate this, 
Figure~\ref{fig:per-num} shows a histogram of periods for 15 BS
binaries formed during 1+3 encounters via these two generic merger
scenarios.  In order to obtain this plot, we have 
used the observed period-eccentricity distribution for the regular
MS-MS binary population in NGC 188 from \citet{geller09} to obtain
estimates for the orbital energies of any binaries and/or triples
going into encounters.  Specifically, in order to obtain periods for
the outer orbits of triples undergoing encounters, we randomly sampled 
the regular period distribution, including only those 
binaries with periods satisfying $400$ days $< P < 4000$ days.  We
have shown that the initial outer orbits of triples going into 1+3
encounters provide a rough lower limit for the periods of BS binaries
formed via Mechanism II.  Therefore, any BS binaries formed in this
way could only have been identified as binaries by radial velocity
surveys if the triple going into the encounter had a period $< 4000$
days (since this corresponds to the current cut-off for detection).
All triples are taken to have a 
ratio of 30 between their inner and outer orbital semi-major axes.
This ratio has been chosen to be arbitrarily large enough that the
triples should be dynamically stable, however we will return to this
assumption in Section~\ref{discussion}.  

We will adopt a ratio based on the observations of
\citet{mathieu09} for the fraction of outcomes that result in each of
these two possible merger scenarios.  In particular, if we assume that
the three BS 
binaries with $P < 150$ days were formed via Mechanism I whereas the
other 12 were formed via Mechanism II (either IIa or IIb since energy
conservation predicts similar periods for the left-over BS binaries),
this would suggest that Mechanism II is $\sim 4$ times more likely to
occur than Mechanism I during any given 1+3 encounter.  We will return
to this assumption in Section~\ref{discussion}.

\begin{figure}
\begin{center}
\includegraphics[width=\columnwidth]{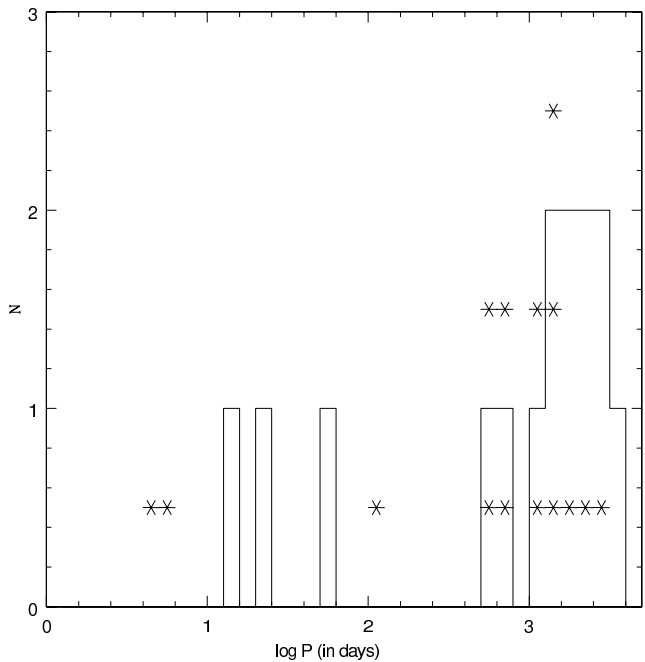}
\end{center}
\caption[Histogram of the period distribution expected for BS binaries
formed during 1+3 encounters]{Histogram of the period distribution (in
  days) expected for BS binaries formed during 15 1+3 encounters.
  The parameter space assumed for the encounters is described in the
  text.  The stars show the observed BS binary period
  distribution in NGC 188 taken from \citet{mathieu09}, where each
  star represents a single BS binary.
\label{fig:per-num}}
\end{figure}

As a result of the requirement for conservation of angular
momentum, we would expect to see a large spread in the distribution of
eccentricities for BSs in long-period binaries formed from encounters
involving triples.  This is because conservation of momentum requires
that the total momentum going into the encounter must be
equal to the total momentum left over after the interaction.  
However, the total initial momentum depends not only on the
initial orbital eccentricities of any binaries or triples going into the
encounter, but also the relative orientations and trajectories of the
colliding objects.  Since the relative orientations and trajectories
are random, the final eccentricities of the BS binaries can take on a
range of values.  In other words, we cannot predict the final distribution of
momenta for BS binaries formed from dynamical encounters.  However,
the observed BS binaries in NGC 188 are observed to 
have a range of eccentricities and this is not inconsistent with a
dynamical origin involving triples.

BSs in short-period binaries formed via Mechanism I can also end up
with just about any eccentricity immediately after the encounter for
the same reasons outlined above.  However, tidal effects become increasingly significant
with decreasing orbital separation so that the hardest binaries should
typically circularize the fastest.  Although theoretical estimates for
the rate of tidal circularization are uncertain \citep{meibom05},
the circularization cut-off period is estimated to be $\sim$ 15 days in NGC
188 \citep{mathieu04}.  From this, it is entirely possible that
recently formed BS 
binaries with P $\sim$ 10 days have not yet become fully
circularized.  

\end{enumerate}

\section{Summary \& Discussion} \label{discussion}

In this paper, we have presented a generalized analytic prescription
for energy conservation during stellar encounters.  Our method can be
used to identify the most probable
dynamical formation scenario for an observed binary or triple
system containing one or more merger products.  We have shown
that, using the observed 
orbital parameters of the system, the allowed initial orbital
semi-major axes of any binary or triple
systems involved in its formation can be constrained.  The
initial semi-major axes of the orbits in turn provide an 
estimate for the collisional cross section and therefore the time-scale
for the encounter to occur in its host cluster.  In order to apply our
technique, repeated spectroscopic measurements of the binary or triple
system containing the merger product(s) are needed in order to obtain
its orbital solution and systemic velocity.  
However, the time-scales provided in Appendix~\ref{appendix} can still
be applied if only the fraction of binary and triple stars are known, 
which can be determined either spectroscopically
\citep[e.g.][]{mathieu90, latham05} or photometrically
\citep[e.g.][]{fan96}.

As we have shown, consideration of the requirement for energy
conservation is ideal for identifying trends during
stellar encounters, whereas 
numerical scattering experiments can require hundreds or even thousands of
simulation runs before patterns will emerge.  Some of these trends include: 

\begin{itemize}

\item At least one short-period binary is usually required in a
  dynamical interaction to produce another binary having a similarly
  short-period (provided no stars are ejected with escape velocities
  $\gtrsim 100$ km/s).  This is because the orbital energy of a
  short-period binary is sufficiently negative that it tends to
  considerably outweigh the other energy terms in
  Equation~\ref{eqn:energy-conserv} for most of the encounters that
  typically occur in the cores of globular and especially open
  clusters.  This has been confirmed by \citet{hurley05}. 

\item Previous studies have found that in order for triples to remain
  stable for many dynamical times, the ratio of their inner to outer
  orbital periods must be relatively large (roughly a factor of ten or
  more) \citep[e.g.][]{mardling01}.  Based on our results, this has
  two important corollaries for stellar mergers in dense cluster
  environments hosting a significant population of triples:

\begin{enumerate}

\item  The longest-lived triples will contain very hard inner binaries
with a large $|\epsilon|$.  This is important since stellar radii are 
in general more likely to overlap and hence mergers to occur
during resonant interactions involving very hard binaries
\citep[e.g.][]{fregeau04, hurley05}.  We therefore expect stellar mergers
to be common during encounters involving stable triple systems.

\item  The longest-lived triples will contain wide outer orbits, creating
a large cross section for collision.  This suggests that the 
time-scale required for a stable triple system to encounter another
object is typically short relative to the cluster age in dense
environments.  A significant fraction of encounters
involving very hard binaries, and hence resulting in stellar mergers,
will therefore involve triples in old open clusters such as M67
and NGC 188. 

\end{enumerate}

\end{itemize}

\citet{vandenberg01} and \citet{sandquist03} suggest that
back-to-back binary-binary encounters, or even a single 3+3 encounter,
could have formed S1082.  We have
improved upon these previous studies by estimating time-scales required
for possible dynamical formation scenarios to occur.  
Since we have argued in Section~\ref{results} that the
formation of S1082 must have involved at least 6 stars, it follows that 
only a 3+3 interaction could have reproduced the observed configuration
via a single encounter.  However, the derived 3+3 encounter time-scale
is sufficiently long that we expect very few, if any, 3+3 encounters
occurred within the lifetime of a typical merger product.  Moreover,
we have argued that the times for multiple encounters to occur are
longer than the cluster age.  Although we cannot rule out a dynamical
origin for S1082, our results suggest that it is unlikely (provided the
derived encounter time-scales were not significantly higher in the
recent past, which we will return to below).  From this, it follows that a
dynamical link between the close binary and third star is unlikely to
exist.   

On the other hand, we have so far ignored the cluster evolution,
and assumed that the currently observed cluster parameters have not
changed in the last few Gyrs.  N-body models suggest that the central
density in M67 could have been significantly higher in the recent
past.  Specifically, Figure 5 of \citet{hurley05} indicates that the
presently observed central density in M67 could have been higher
within the past Gyr by a factor of $\gtrsim 2$.  If this was indeed
the case, our previous estimates for each of the different encounter
frequencies should increase by a factor of $\sim 4$, so that a
significant number of dynamical encounters involving single, binary
and triple stars should have occurred in M67 within the last
$\tau_{BS}$ years.  It follows that a dynamical
origin for S1082 is not unlikely if the central density in M67 was 
recently larger than its presently observed value by a factor of
$\gtrsim 2$.  This also increases the 
probability that a scenario involving multiple encounters created
S1082, although we have shown that such a scenario is still likely to
have involved one or more triples.  

Based on the preceding arguments, S1082 offers an excellent example of
how observations of 
individual multiple star systems containing BSs can be used to
directly constrain the dynamical history of their host cluster.  
If a definitive dynamical link between components A and B is 
established, this would suggest that the central density in M67 
was higher in the last 1-2 Gyrs.  This is also required in order
for the cluster dynamics to have played a role in the formation of a
significant fraction of the observed BS population in M67.  Based on
the current density, the encounter time-scales are sufficiently long
that too few encounters should have occurred in the last $\tau_{BS}$
years for mergers during dynamical interactions to be a significant
contributor to BS formation.

We have obtained quantitative constraints 
for two generic channels for mergers during encounters involving
triples -- one in which a direct stellar collision occurs within a
dynamical interaction of the hard inner binary of a triple and another
single or (hard) binary star (Mechanism I) and
one in which the hard inner binary of a triple is driven to coalesce
by imparting energy and/or angular momentum to other stars involved in the
interaction (Mechanism II).  Our results
suggest that these two general merger mechanisms could contribute to a
bi-modal period distribution for BS binaries similar to that
observed in NGC 188.  These dual mechanisms predict a gap in
period, with those BS binaries formed via Mechanism I having periods of a
few to $\sim 100$ days and those formed via Mechanism II having
periods closer to $\sim 1000$ days.  Some 2+2, 2+3 and even 1+3
encounters could involve orbits with periods in this range, and
Equation~\ref{eqn:energy-conserv2} confirms that the final period of
a BS binary formed via Mechanism II will typically be determined by
that of the second hardest binary orbit.  Therefore, we might still
expect some BS binaries to have periods that fall in the gap (100 days
$\lesssim$ P $\lesssim$ 1000 days).  Our results do indeed predict one
such BS binary, as shown in Figure~\ref{fig:per-num}.  

A number of assumptions went into obtaining Figure~\ref{fig:per-num},
many of which were chosen specifically to reproduce the observed BS
binary period distribution.  Regardless, our
assumptions were chosen to reflect encounter scenarios that are the
most likely to result in mergers.  These should involve triples with
very hard inner binaries since these are the most likely to merge
during encounters.  The triples should also have outer companions on
very wide orbits since these have the largest cross sections for
collision.  From this, we have assumed a ratio of
30 between the inner and outer semi-major axes of all triples.  This
ensures that all triples are dynamically stable.  This also leads us
to assume a minimum period of $400$ days for the outer
orbits of triples so that the corresponding minimum period of their
inner orbits is not too small.  These assumptions serve to show that
encounters involving 
triples could produce both long-period and short-period BS binaries as
well as a period gap.  

Our results predict that the short-period peak in
Figure~\ref{fig:per-num} is at a slightly longer period than the
observations suggest.  If we decrease the assumed  
ratio between the inner and outer orbital separations of triples, the
short-period peak will move to even longer periods.  However, if one
or more stars were ejected with a high escape velocity or the
dissipative effects of tides are considered (which are expected to be
the most significant for encounters involving hard binaries) this
would move the short-period peak to even shorter periods.  In order to
obtain the desired agreement with the observations at the short-period
end of the BS period distribution, energy must have somehow been
dissipated or removed from the hard inner binaries of the triples
during (or even after) the encounter, or
the inner binaries must have initially been even harder than 
we have assumed in obtaining Figure~\ref{fig:per-num}.  

Our results could suggest that the hard BS binaries in
NGC 188 (binaries 5078 and 7782) may have outer triple companions,
perhaps with sufficiently long periods that they would have thus far
evaded detection.  This is consistent with the requirement for energy
conservation since the orbital energy of the outer
orbit of the triple is negligible compared to that of its inner
binary.  If binaries 5078 and 7782 do have outer triple companions, it
is also possible that Kozai oscillations combined with tidal friction
contributed to decreasing their orbital periods \citep{eggleton06}.
Finally, the BS binaries in NGC 188 are observed 
to have a wide range of eccentricities, which we have argued is
not inconsistent with a dynamical origin involving triples.  

We have assumed that Mechanism II is more likely to occur than
Mechanism I during any given 1+3 encounter.  This is a
reasonable assumption since numerical scattering experiments of 1+2
and 2+2 encounters have shown that the coalescence
of a hard binary is much more likely to occur during an encounter than
a direct collision between a hard binary and an interacting single
or binary star \citep[e.g.][]{fregeau04}.  The ratio we have
assumed for the frequencies with which Mechanisms I and II occur was
specifically chosen in order to reproduce the observed numbers of
short- and
long-period BS binaries.  The important point to take away is that the
observed BS 
binary period-eccentricity distribution offers a potential constraint
on the fraction of encounters that result in different merger
scenarios.  

Based on our results, Mechanism II must occur $\sim 4$ times more
often than Mechanism I in order to reproduce the observed BS period
distribution from 1+3 encounters (or, equivalently, 2+3 encounters
involving a very wide binary and 2+2 encounters between a 
short-period binary and a long-period binary).  This can be tested by performing
numerical scattering experiments of encounters involving triples.
Therefore, our
results highlight the need for simulations of
1+3, 2+3 and 3+3 encounters to be performed in order to better
understand their expected contributions to BS populations in open and
globular clusters.  Once a preferred encounter scenario has been
identified for an observed binary or triple containing one or more
BSs, numerical scattering experiments can be used to
further constrain the conditions under which that scenario will occur
(or to show 
that it cannot occur).  We have demonstrated that a combination of 
observational and analytic constraints can be used to isolate the
parameter space relevant to the dynamical formation of an observed
multiple star system (or population of star systems) containing
one or more merger products.  This will drastically narrow the
relevant parameter space for numerical scattering experiments.

We have improved upon the results of \citet{perets09} and
\citet{mathieu09} since we have shown that dynamical encounters
involving triples could not only be contributing to
the long-period BS binaries in NGC 188, but they could also be an
important formation
mechanism for short-period BS binaries and triples containing BSs.  We
have not ruled out mass transfer or Kozai-induced 
mergers in triples (primordial or otherwise) \citep{mathieu09,
  perets09}, or even various 
combinations of different mechanisms, as contributing formation channels
to the BS binary population in NGC 188.  For instance, a 1+3 exchange
interaction could stimulate a merger indirectly if the resulting angle
of inclination between the inner and 
outer orbits of the triple exceeds $\sim 39^{\circ}$, ultimately
allowing
the triple to evolve via the Kozai mechanism so that the eccentricity
of the inner binary increases while its period remains roughly
constant \citep{eggleton06}.  

There is evidence to suggest that mass transfer via Roche lobe
over flow could play a
role in the formation of at least some BSs.  It is difficult to account for 
the near zero eccentricities of some of the long-period BS binaries
without at least one episode of mass transfer having occurred. This is
because none of the normal MS-MS binaries with similar periods have
such small eccentricities \citep{mathieu09}.  On the other hand, it
may not be unreasonable to expect that some collision products left in
binaries undergo mass transfer since they are expected to expand
adiabatically post-collision, and will sooner or later evolve to
ascend the giant branch.  As a result of conservation
of energy and angular momentum, the mass transfer process will usually
act to 
increase the orbital periods of these binaries provided it is
conservative \citep{iben91}.  Interestingly, the cut-off period for
Roche lobe overflow is $\sim 1000$ days for low-mass stars
\citep{eggleton06}, which is in rough agreement with the long-period
peak in the observed period-eccentricity distribution of the BS binary
population in NGC 188.  Therefore, mass transfer could also be
contributing to the 
period gap observed for the BS binaries.  

According to the results of
\citet{geller09}, the number of giant-MS binaries with $P \lesssim
1000$ days is comparable to the
number of BS binaries (A. Geller, private communication).  It is unlikely that
every giant-MS binary will form a BS from mass transfer, however,
suggesting that at most a few of the long-period BS binaries in NGC
188 were formed via this mechanism.  Finally, if the outer
companion of a triple system evolves to
over-fill its Roche lobe it could transfer mass to both of the
components of the close inner binary.  This mechanism could therefore
also produce two BSs in a close binary, although it predicts the presence of
an orbiting triple companion.  For these reasons,
a better understanding of triple evolution, as well as binary
evolution in binaries containing merger products, is needed.

The dissipational effects of tides tend
to convert stars' bulk translational kinetic energies into internal or
thermal energy within the stars, leading to an increase in the total
gravitational binding energy of the stellar configuration
\citep[e.g.][]{mcmillan90}.  By
increasing the terms U$_{ii}$ in Equation~\ref{eqn:energy-conserv},
the initial orbital energies of any binaries going into an encounter
can increase accordingly in order to conserve energy.  A higher orbital energy
corresponds to a larger semi-major axis and hence cross section for
collision.  This suggests that the derived encounter time-scales can be
taken as upper limits in the limit that tidal dissipation is
negligible.  We expect tides to be particularly
effective during encounters for which the total energy is very
negative as a result of
one or more very hard binaries being involved.  

We have argued in Section~\ref{general} that the average stellar 
mass is expected to be comparable to (but slightly less than) the mass
of the MSTO in old OCs and 
low-mass GCs.  We have also argued that most encounters will involve
stars having masses slightly larger than the average stellar mass.  We
might therefore expect to 
find that a high proportion of merger products have masses that
exceed that of the MSTO in very dynamically-evolved clusters that have
lost a large fraction of their low-mass stars.  Consequently, a larger 
number of merger products could appear sufficiently bright to
end up in the BS region of the cluster CMD in these clusters than in
their less dynamically-evolved counterparts.  This 
is consistent with the results of \citet{knigge09} and
\citet{leigh09} 
who found that the number of BSs in the cores of GCs scales
sub-linearly with the core mass.  In particular, since the cluster
relaxation time 
increases with increasing cluster mass, it is the lowest mass GCs that
should have lost the largest fraction of their low-mass stars.
Therefore, if a larger fraction of merger products do indeed end up
more massive than the MSTO in these clusters, this could be a
contributing factor to the observed sub-linear dependence on core
mass.  It is also interesting to note that, since BSs are among the
most massive cluster members and many are thought to have a binary
companion, BSs should be preferentially retained in clusters as they
evolve dynamically compared to low-mass MS stars.  This could also
contribute to the observed sub-linear dependence on core mass.  

The exchange or conversion of energies that occurs during an
encounter takes place over a finite period of time, so it is important to
specify whether or not the system has fully relaxed
post-encounter when discussing the remaining stellar configuration.
For one thing, \citet{sills01} showed that, although collision
products may be in hydrodynamic equilibrium, they are not in thermal
equilibrium upon formation and so contract on a thermal time-scale.
Simulations also suggest that most merger products should be rapid
rotators \citep{sills02, sills05}.  However, at least in the case of
blue stragglers, this is
rarely supported by the observations.  Some mechanism for
angular momentum loss must therefore be operating either during or
after the merger takes place in order to spin down the remnant.
The time-scale
considered must also be sufficiently short that subsequent dynamical
interactions are unlikely to have occurred since these could affect
the total energy and momentum of the system.  

With this last point in mind, N-body simulations considering BS
formation have shown that after they are formed, BSs are often
exchanged into other multiple star systems \citep{hurley05}.  This
suggests that for multiple star systems containing more than one BS,
the BSs could have first been formed separately or in parallel, and
then exchanged into their presently observed configuration.  For the
case of S1082, this would require at least 3 separate dynamical
interactions.  Given that the derived times between encounters are
relatively long and the fact
that the most likely formation scenario is usually that for which the
number of encounters is minimized, the current state of M67 suggests
that the probability of S1082 having formed from a scenario involving
3 encounters is low.  Conversely, the derived encounter time-scales in
NGC 188 are sufficiently short that many of the BS binaries 
could have experienced a subsequent dynamical interaction after their
formation.  BSs tend to 
be more massive than normal MS stars, contributing to an increase in
their gravitationally-focussed cross section for collision.  This
suggests that the encounter time-scale for multiple star systems
containing BSs is slightly shorter than for otherwise
identical systems composed only of normal MS stars.  This contributes
to a slight increase in the probability that a BS will experience an
exchange encounter after it is formed.  Interestingly, it could also contribute to
an increase in the probability that a close binary containing two BSs
will form during an encounter between two different multiple star
systems each containing their own BSs (\citet{mathieu09}; R. Mathieu,
private communication).  This is because it is the
heaviest stars that will experience the strongest gravitational
focussing and are therefore the most likely to experience a close
encounter, end up in a closely bound configuration, or even merge.

Most exchange interactions will involve wide binaries for which
the cross section for collision 
is large.  Since wide binaries are typically relatively soft and the
hardest binary involved in the interaction will usually determine the
orbital energy of the left-over BS binary, most exchange interactions
will leave the periods of BS binaries relatively unaffected.  This
need not be the case, of course, provided one or more stars are
ejected from the system with a very high escape velocity.  With
these last points in mind, we have assumed throughout our analysis that
all binaries and triples are dynamically hard.  This is a reasonable
assumption since the hard-soft boundary corresponds to a period of
$\sim 10^6$ days in both M67 and NGC 188, for which the cross section
for collision is sufficiently large that we do not expect such
binaries to survive for very long.  Nonetheless, considerations such
as these must be properly taken into account when isolating a
preferred formation scenario and predicting the final distribution of
energies.

We have presented an analytic technique to constrain the dynamical
origins of multiple star systems containing one or more BSs.  Our
results suggest that, in old open clusters, most dynamical
interactions resulting in mergers involve triple stars.  If most
triples are formed dynamically, this could suggest that
many stellar mergers are the culmination of a hierarchical build-up of
dynamical interactions.  
Consequently, this mechanism for BS formation should 
be properly included in future N-body simulations of cluster
evolution.  A better
understanding of the interplay between the cluster dynamics and the
internal evolution of triple systems is needed in order to better
understand the expected period distribution of BS binaries formed from
triples.  Simulations will therefore need to
track both the formation 
and destruction of triples as well as their internal evolution via
Kozai cycles, stellar and binary evolution, etc.  On the observational
front, our results highlight the need for a more detailed knowledge of
binary and especially triple populations in clusters. 

\appendix

\section{Collisional Cross Sections and Time-scales} \label{appendix}

The gravitationally-focused cross sections for 1+1,
1+2, 2+2, 1+3, 2+3 and 3+3 collisions can be found using
Equation 6 from \citet{leonard89}.  Neglecting the first term and
assuming that binary and triple stars are on average twice and three
times as massive as single stars, respectively, this gives for the
various collisional cross sections:
\begin{equation}
\label{eqn:cs-1+1}
\sigma_{1+1} \sim \frac{8{\pi}GmR}{v_{rel}^2},
\end{equation}
\begin{equation}
\label{eqn:cs-1+2}
\sigma_{1+2} \sim \frac{3{\pi}Gma_b}{v_{rel}^2},
\end{equation}
\begin{equation}
\label{eqn:cs-2+2}
\sigma_{2+2} \sim \frac{8{\pi}Gma_b}{v_{rel}^2},
\end{equation}
\begin{equation}
\label{eqn:cs-1+3}
\sigma_{1+3} \sim \frac{4{\pi}Gma_t}{v_{rel}^2},
\end{equation}
\begin{equation}
\label{eqn:cs-2+3}
\sigma_{2+3} \sim \frac{5{\pi}Gm(a_b + a_t)}{v_{rel}^2},
\end{equation}
\begin{equation}
\label{eqn:cs-3+3}
\sigma_{3+3} \sim \frac{12{\pi}Gma_t}{v_{rel}^2}.
\end{equation}
Values for the
pericenters assumed for the various types of encounters are shown in
Table~\ref{table:peri}, where $R$ is the average stellar
radius, $a_b$ is the average binary semi-major axis and $a_t$ is the
average semi-major axis of the outer orbits of triples.

\begin{table}
\centering
\caption{Pericenters Assumed for Each Encounter Type
  \label{table:peri}}
\begin{tabular}{lc}
\hline
Encounter Type & Pericenter \\
\hline
1+1 & $2R$ \\
1+2 & $a_b/2$ \\
1+3 & $a_t/2$ \\
2+2 & $a_b$ \\
2+3 & $(a_b + a_t)/2$ \\
3+3 & $a_t$ \\
\hline
\end{tabular}
\end{table}

In general, the time between each of the different encounter
types can be found using Equation~\ref{eqn:coll-rate} and
the gravitationally-focused cross sections for collision given 
above.  Following the derivation of \citet{leonard89}, we can write
the encounter rate in the general form:
\begin{equation}
\label{eqn:coll-rate-gen}
\Gamma_{x+y} = N_xn_y{\sigma}_{x+y}v_{x+y},
\end{equation}
where $N_x$ and $n_y$ are the number and number density, respectively,
of single, binary or triple stars and $v_{x+y}$ is the relative
velocity at infinity between objects $x$ and $y$.  For instance, the
time between binary-binary encounters in the core of a cluster is
given by: 
\begin{equation}
\begin{gathered}
\label{eqn:coll2+2}
\tau_{2+2} = 1.3 \times 10^7f_b^{-2} \Big(\frac{1
  pc}{r_c}\Big)^3\Big(\frac{10^3
  pc^{-3}}{n_0}\Big)^2 \\
\Big(\frac{v_{rms}}{5 km/s}\Big)\Big(\frac{0.5 M_{\odot}}{<m>}\Big)\Big(\frac{1
  AU}{a_{b}}\Big) \mbox{years}.
\end{gathered}
\end{equation}
Similarly, the times between 1+1, 1+2, 1+3, 2+3 and 3+3 encounters are
given by:
\begin{equation}
\begin{gathered}
\label{eqn:coll1+1}
\tau_{1+1} = 1.1 \times 10^{10}(1-f_b-f_t)^{-2}\Big(\frac{1 pc}{r_c}
\Big)^3 \Big(\frac{10^3 pc^{-3}}{n_0} \Big)^2 \\
\Big(\frac{v_{rms}}{5 km/s} \Big) \Big(\frac{0.5 M_{\odot}}{<m>} \Big)
\Big(\frac{0.5 R_{\odot}}{<R>} \Big)\mbox{ years},
\end{gathered}
\end{equation}
\begin{equation}
\begin{gathered}
\label{eqn:coll1+2}
\tau_{1+2} = 3.4 \times 10^7(1-f_b-f_t)^{-1}f_b^{-1} \Big(\frac{1 pc}{r_c}
\Big)^3 \Big(\frac{10^3 pc^{-3}}{n_0} \Big)^2 \\
\Big(\frac{v_{rms}}{5
  km/s} \Big) \Big(\frac{0.5 M_{\odot}}{<m>} \Big) \Big(\frac{1
  AU}{a_{b}} \Big)\mbox{ years},
\end{gathered}
\end{equation}
\begin{equation}
\begin{gathered}
\label{eqn:coll1+3}
\tau_{1+3} = 2.6 \times 10^7(1-f_b-f_t)^{-1}f_t^{-1} \Big(\frac{1 pc}{r_c}
\Big)^3 \Big(\frac{10^3 pc^{-3}}{n_0} \Big)^2 \\
\Big(\frac{v_{rms}}{5
  km/s} \Big) \Big(\frac{0.5 M_{\odot}}{<m>} \Big) \Big(\frac{1
  AU}{a_{t}} \Big)\mbox{ years},
\end{gathered}
\end{equation}
\begin{equation}
\begin{gathered}
\label{eqn:coll2+3}
\tau_{2+3} = 2.0 \times 10^7f_b^{-1}f_t^{-1} \Big(\frac{1 pc}{r_c}
\Big)^3 \Big(\frac{10^3 pc^{-3}}{n_0} \Big)^2 \\
\Big(\frac{v_{rms}}{5
  km/s} \Big) \Big(\frac{0.5 M_{\odot}}{<m>} \Big) \Big(\frac{1
  AU}{a_{b}+a_{t}} \Big)\mbox{ years},
\end{gathered}
\end{equation}
and
\begin{equation}
\begin{gathered}
\label{eqn:coll3+3}
\tau_{3+3} = 8.3 \times 10^6f_t^{-2} \Big(\frac{1 pc}{r_c}
\Big)^3 \Big(\frac{10^3 pc^{-3}}{n_0} \Big)^2 \\
\Big(\frac{v_{rms}}{5
  km/s} \Big) \Big(\frac{0.5 M_{\odot}}{<m>} \Big) \Big(\frac{1
  AU}{a_{t}} \Big)\mbox{ years}.
\end{gathered}
\end{equation}

\section*{Acknowledgments}

We would like to thank Bob Mathieu for many helpful comments and
suggestions.  We would also like to thank Aaron Geller, Evert
Glebbeek, Hagai Perets, David Latham, Daniel Fabrycky and Maureen van
den Berg for useful discussions.  This research has been supported by
NSERC as well as the National Science Foundation under Grant
No. PHY05-51164 to the Kavli Institute for Theoretical Physics.

\bsp

\label{lastpage}

\end{document}